\documentclass[10pt,letterpaper]{article}
\usepackage[top=0.85in,left=2.75in,footskip=0.75in]{geometry}

\usepackage{amsmath,amssymb}

\usepackage{changepage}

\usepackage{textcomp,marvosym}

\usepackage{cite}

\usepackage{nameref,hyperref}

\usepackage{float} 
\usepackage[right]{lineno}
\hypersetup{breaklinks=true}

\usepackage[nopatch=eqnum]{microtype}
\DisableLigatures[f]{encoding = *, family = * }

\usepackage[table]{xcolor}

\usepackage{array}

\newcolumntype{+}{!{\vrule width 2pt}}

\newlength\savedwidth

\raggedright
\usepackage[aboveskip=1pt,labelfont=bf,labelsep=period,justification=raggedright,singlelinecheck=off]{caption}

\makeatletter
\renewcommand{\@biblabel}[1]{\quad#1.}
\makeatother

\usepackage{tabularx} 
\usepackage{booktabs} 
\usepackage{setspace} 
\usepackage{placeins} 

\usepackage{lastpage,fancyhdr,graphicx}
\usepackage{epstopdf}
\fancyheadoffset[L]{2.25in}
\fancyfootoffset[L]{2.25in}
\begin{document}
\vspace*{0.2in}

\begin{flushleft}
{\Large
\textbf\newline{Digital Epidemiology after COVID-19: impact and prospects} 
}
\newline
\\
Sara Mesquita\textsuperscript{1,2},
Lília Perfeito\textsuperscript{1},
Daniela Paolotti\textsuperscript{4},
Joana Gonçalves-Sá\textsuperscript{1,3*},
\\
\bigskip
\textbf{1} LIP -- Laboratory for Instrumentation and Experimental Particle Physics, Avenida Prof. Gama Pinto 2, 1600-078 Lisboa, Portugal
\\
\textbf{2} Nova Medical School, Campo Mártires da Pátria 130, 1169-056 Lisboa, Portugal
\\
\textbf{3} Nova School of Business and Economics, Rua da Holanda, 2775-405 Carcavelos, Portugal
\\
\textbf{4} ISI Foundation, via Chisola 5, 10126 Turin, Italy
\\
\bigskip

%
%





* joanagsa@lip.pt

\end{flushleft}
\section*{Abstract}
Epidemiology and Public Health have increasingly relied on structured and unstructured data, collected inside and outside of typical health systems, to study, identify, and mitigate diseases at the population level. Focusing on infectious disease, we review how Digital Epidemiology (DE) was at the beginning of 2020 and how it was changed by the COVID-19 pandemic, in both nature and breadth. We argue that DE will become a progressively useful tool as long as its potential is recognized and its risks are minimized. Therefore, we expand on the current views and present a new definition of DE that, by highlighting the statistical nature of the datasets, helps in identifying possible biases. We offer some recommendations to reduce inequity and threats to privacy and argue in favour of complex multidisciplinary approaches to tackling infectious diseases.

\section*{Introduction}

Epidemiology is the study of health patterns and determinants in a population~\cite{cdcepi}. It relies on diverse health-related data from various sources, including questionnaires, laboratory tests, and socio-demographic information. Recent years have witnessed the rise of Digital Epidemiology (DE), a sub-field of Epidemiology, fueled by the pervasive adoption of digital technology (data availability) and computational power~\cite{aiello2020social}. It was originally defined as the use of digital data collected for non-epidemiological purposes in epidemiological studies~\cite{salathe2012digital}, and the COVID-19 pandemic greatly increased its breadth, both in terms of data sources and their applications.

However, the current DE definition~\cite{salathe2018digital} falls short for three primary reasons: first, datasets are becoming universally digital, encompassing clinical, social network, and classical field survey data; second, epidemiology increasingly relies on non-traditional but clinical datasets like electronic prescription records and on-call triage systems for disease incidence estimation and syndromic surveillance; and third, epidemiology has a long history of re-purposing datasets, including those related to public housing, human and animal density, traffic, weather, or postal codes, far extending beyond data collected solely for epidemiological studies.

Thus, we find that the main difference between traditional (or classical - CE) and "Digital Epidemiology" (DE) datasets, is not so much the original purpose or the type of data storage, but how the collection is designed and the results generalized from the initial samples. In particular, and regardless of the data used, while CE studies are typically created and designed for a specific public health-related purpose, DE datasets are usually collected without worrying about representativeness and in a complementary fashion, and this creates an important statistical distinction with two main implications for bias correction: 1) Traditional datasets are carefully collected \textit{a priori} to ensure statistical representativeness, while biases in DE datasets are addressed \textit{a posteriori}; 2) DE datasets are often tapping different  (sub-)populations, and are sometimes collected without explicit consent.
Therefore, due to different strengths and weaknesses, these two approaches are strongly complementary and our proposed definition of Digital Epidemiology, emphasizing the statistical nature of the data, helps identify areas for improvement and utility.

We focus on infectious diseases, as these comprise over 50\% of DE articles~\cite{park2018digital}. This is not surprising as established surveillance systems typically rely on processes that are laborious, costly, and slow, such as sentinel doctors, surveyors, and testing labs~\cite{velasco2018disease}. DE complements these established surveillance systems by utilizing data streams from various sources, enabling rapid identification of disease dynamics, risk factors, and optimized response strategies~\cite{budd2020digital}. In particular, the COVID-19 pandemic highlighted DE's potential, leading to significant transformations~\cite{milne2020disruption}.

We begin with a summary of early 2020's DE of infectious diseases and explore pandemic-induced changes in systems (data sources, approaches, etc.). Subsequently, we provide recommendations to enhance DE's progressive utility, emphasizing the impact of COVID-19 and considering statistical and analytical perspectives.

\section*{Digital Epidemiology before the COVID-19 Pandemic}

DE's potential to enhance surveillance, making it faster, cheaper, and broader, is not new. A notable example is the 1984 creation of a network of sentinel doctor digital records, in France, focusing on influenza syndromic monitoring~\cite{valleron1986computer}. In fact, pre-COVID-19, flu was frequently studied in DE, highlighting both its potential and limitations, and we offer some examples below.

Launched in 2008, Google Flu Trends (GFT) aimed to correlate online searches for flu-like symptoms with cases~\cite{ginsberg2009detecting}. While initially successful, GFT missed the 2009 pandemic onset and overestimated cases by 140\% during the 2012–13 flu season~\cite{lazer2014parable}. It has been argued that its accuracy was affected by spurious correlations, over-fitting, and media coverage~\cite{olson2013reassessing}, and this might also be a problem when analyzing social media posts~\cite{tizzoni2020impact}. 

Still, the first 2009 flu pandemic cases were reported by local news media in Mexico before being detected by the CDC or the WHO, supporting use of non-traditional data and in different languages~\cite{freifeld2008healthmap}. Indeed, in 2013, the CDC announced FluSight, a competition to improve flu monitoring, and the winning methodology used a combination of data sources, including weather and online searches~\cite{influenza_challenge, santillana2015combining}. 

Other efforts have been leveraging online platforms to include citizen participation and self-reporting. For instance, InfluenzaNet collects voluntary flu-like symptoms from people across 10 European countries, having been monitoring influenza for over a decade, often anticipating official records~\cite{koppeschaar2017influenzanet}, and similar successful approaches were adopted in other countries~\cite{smolinski2015flu, moberley2019flutracking, lwin2017flumob}. While self-reporting provides a cost-effective way to gather longitudinal data and engages participants transparently and ethically, it is subject to several biases and implemented models must correct for them (see Table~\ref{tab:Table 1}). 

In the early 2000's several countries also started free or low-cost phone services to offer clinical advice and triage callers. Having patient data and symptoms as inputs, these systems combine pre-determined computational algorithms with curation by trained clinicians. Covering increasingly large demographics, they can be used for large-scale surveillance and showed potential for fast outbreak detection~\cite{won2017early}.

Tracing tools based on cell-phones, WiFi, or Bluetooth badges~\cite{yoneki2011fluphone, danquah2019use, farrahi2014epidemic} revolutionized spatial disease models, aiding public health interventions during outbreaks of Ebola~\cite{vorovchenko2017ebola}, Dengue~\cite{albinati2017enhancement}, Zika~\cite{mcgough2017forecasting}, and others~\cite{de2022intelligent}. However, such studies tend to oversample from well-off populations~\cite{hargittai2020potential}, and just increasing cell-phone ownership does not guarantee much-needed digital literacy and access. Other approaches take advantage of airline traffic, commuter trajectories~\cite{ charaudeau2014commuter}, general mobility~\cite{tizzoni2014use}, or urban lights to explain disease transmission~\cite{bharti2011explaining}.

Overall, DE studies on communicable diseases steadily increased from 2005 to January 2020, leveraging text data from social media (Twitter, Facebook, and Instagram), search engines (Google and Baidu), news media, and Wikipedia. Goals encompassed disease prediction, exploring seasonal dynamics, information-seeking behavior, and detecting outbreaks or pandemics, with flu central to research. Studies addressed both seasonal (33\%) and epidemic (45\%) occurrences, showcasing applicable tools for predicting transmission pattern changes~\cite{shakeri2021digital}.

However, scarce evidence shows widespread adoption of results by public health officials due to lack of validation, funds, or perceived usefulness~\cite{shakeri2021digital}. Rare examples include~\cite{paolotti2014web} and~\cite{neto2020participatory}, which emphasize the significance of integrating traditional and non-traditional data sources.
Other pioneers include the Program for Monitoring Emerging Diseases (ProMED)~\cite{madoff2005internet}, the Global Public Health Intelligence Network (GPHIN)~\cite{blench2007global} or HealthMap~\cite{tarkoma2020fighting}. Also, before 2020, WHO and independent initiatives routinely tracked various diseases (e.g., Zika fever, Ebola, dengue) and identified around 3,000 potential outbreak signals monthly~\cite{sridhar2020covid}.

In summary, by early 2020, these systems showed promise but lacked sufficient exploration. GPHIN, credited with early detection of MERS, Ebola, and the 2009 flu pandemic, was halted in 2019~\cite{globeandmail2023} and, as a result, none of these tools could prevent the spread of SARS-CoV-2, leading to the second pandemic of this century.
\vspace{12pt}

\section*{The impact of the COVID-19 Pandemic}
~\label{3}

At the beginning of 2020, a surge in hospitalizations and deaths was traced to infections from a new coronavirus and led to strong and worldwide government-enforced containment measures.

Great efforts were put into data collection to track infections and infer general patterns and dynamics, at different levels: countries were expected to produce detailed case reports, large platforms united to share data and create digital contact tracing apps, and the lockdowns forced the population to communicate more online, making large amounts of medically relevant data available.

\begin{figure}[H]
\centering
\includegraphics[width=\linewidth]{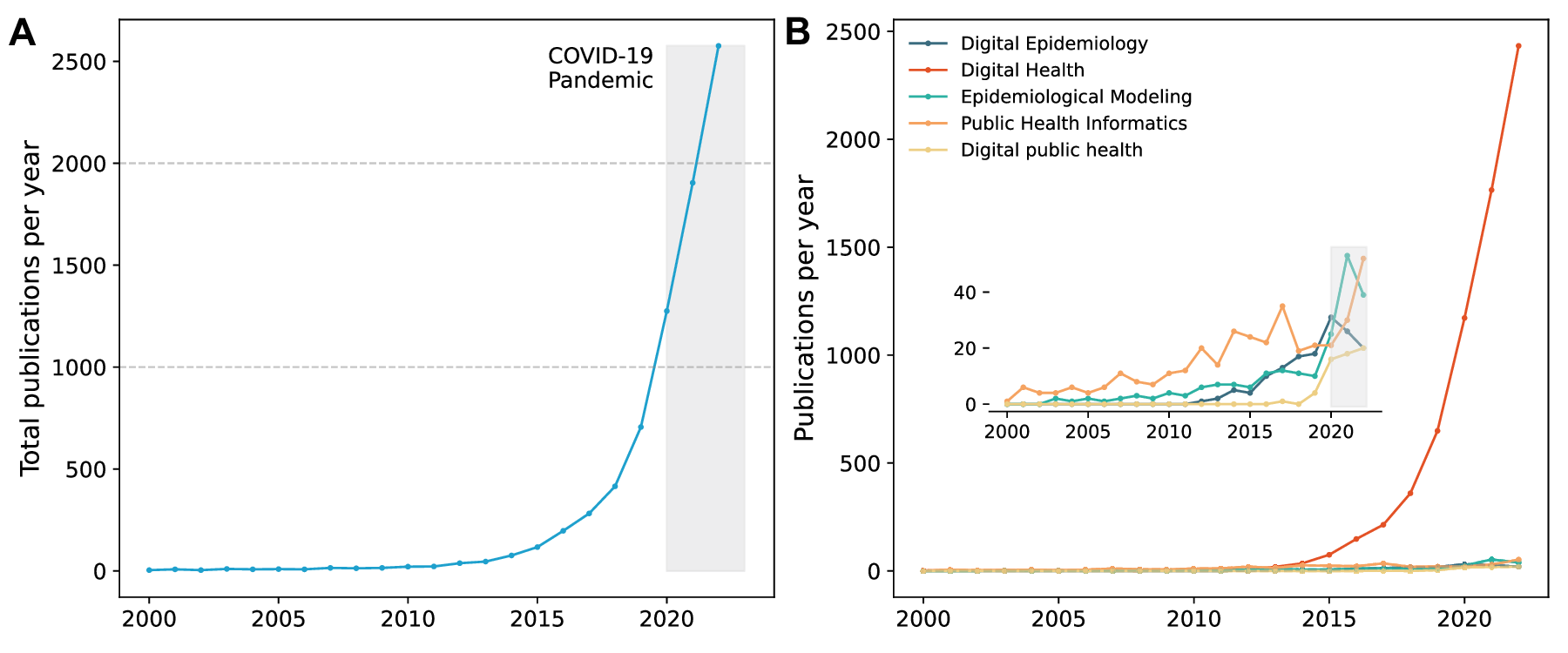}
\caption{\textbf{Digital Epidemiology publications over time.} \textbf{A}, The yearly total number of digital epidemiology publications obtained through a systematic data extraction process using the PubMed API. A comprehensive set of queries was applied to PubMed, including relevant keywords such as 'Digital Epidemiology', 'Digital Health', 'Epidemiological Modeling', 'Public Health Informatics', and 'Digital public health. The resulting data provides insights into the overall trends in digital epidemiology research. \textbf{B}, Publication trends within specific domains related to digital epidemiology. The data spans from 2000 to 2022, providing an overview of how research in these domains has evolved over time.
}
\label{fig:1}
\end{figure}

The pandemic, as depicted in Figure~\ref{fig:1}, brought about a rapid transformation in Digital Epidemiology (DE), characterized by 1) an unprecedented increase in both the quantity and quality of data, 2) the emergence of a new data-sharing culture, and 3) a strong push towards the acceptance of technological solutions, as discussed in~\cite{milne2020disruption}. Overall, we argue that the pandemic accelerated a previously gradual process and fundamentally altered the nature of DE.

Indeed, despite of their major advantages, many of these datasets were not statistically representative and had important biases that could not be accounted for or corrected \textit{a priori}. We list some key ones and discuss their limitations.

The data made available by the Johns Hopkins University Center for Systems Science and Engineering (JHU CSSE)~\cite{dong2022johns}, often included details on positive cases (age, gender), mortality, ICU occupancy, etc., and became a key resource, allowing for model development and predictions. However, the lack of standardized testing reduced the reliability of official numbers and hindered between-country comparisons, particularly when national policies varied. For example, decisions on whether tests were required to stay at home or if self-tests were state-sponsored varied between countries and with time. Similarly, criteria for classifying deaths as caused by or associated with COVID-19 differed~\cite{singh2021international}. 
Thus, large-scale testing was fundamental but comparisons between positivity rates or incidence did not become necessarily easier, as data increased.

With people shifting communication to the digital world and sharing health status, symptoms, and test results~\cite{sarker2020self}, cases and social media reports became closely linked~\cite{alshaabi2021world}. 

While these data over-sampled from internet-savvy populations, it proved useful to back-estimate disease introduction~\cite{menkir2021estimating}, assess compliance with public health measures~\cite{delussu2022evidence}, and gauge vaccination intention~\cite{crupi2022echoes}. In another example, availability of datasets and data visualization software led to social-media-friendly graphs, figures, and dashboards, facilitating communication with the public. However, this instantaneous access could hinder discernment of truth and scientific relevance, influencing trust in science amidst a disinformation pandemic~\cite{gonccalves2020fight}: if such data and tools contributed to early strong compliance~\cite{gupta2021tracking}, they might have later led to increased distrust and vaccine refusal~\cite{de2021relationship}.
\vspace{12pt}

The COVID-19 pandemic also highlighted apps and wearable devices in healthcare, with oxymeter and heart-rate tool sales surging~\cite{vaidheeswaran2021consumer} and digital contact-tracing (DCT) apps gaining visibility~\cite{pandit2022smartphone, ojokoh2022contact}. These utilize proximity sensors (mainly Bluetooth) and exposure notifications: if app users A and B are in close contact and user A tests positive, user B can be alerted to get tested or isolate. This early warning could reduce further exposure and even break transmission chains, but it can be argued that these apps were quite unsuccessful, particularly in countries that did not enforce their use, for several possible reasons. A key concern was privacy, especially after Google and Apple joined forces to create an opt-in smartphone-based system~\cite{sharma2020use, seto2021adoption}. 

Indeed, despite promises that data would be anonymous and private, a major flaw was identified for Android users that jeopardized anonymity and location-sharing was required~\cite{ng2021google}. 
Other trade-offs between efficiency and anonymity proved difficult to balance: no integration with manual contact tracing or only allowing for clinically validated manual insertion of test results offered more privacy-protecting solutions but severely slowed alerts. Finally, countries often did not work closely with medical, behavioral, and communication experts when deploying these tools, to include underprivileged populations or facilitate communication between teams~\cite{bedson2021review}. 
In retrospect, insufficient uptake~\cite{moreno2021anatomy} and lack of user-based design undermined the apps' potential~\cite{cencetti2021digital}: including such large corporations and placing most emphasis on technical solutions might have reduced compliance~\cite{rockhealth}.

Similarly, mobile-based open-source mobility data was made available by Facebook, Google, Apple, and several phone providers. Together with epidemiological data, these datasets proved to be useful in evaluating changes in mobility and compliance with lockdowns~\cite{pullano2020evaluating, pepe2020covid, hazarie2021interplay}, to better understand the evolution of the pathogen~\cite{woskie2021early, lemey2021untangling}, or to minimize disease spread~\cite{kishore2021exploring}. 

While elder and poor communities, with no access to smartphones, were again less visible in these datasets, these studies further revealed differences in exposure~\cite{levy2022neighborhood, woskie2021men}, with people from lower socio-economic status confining less, in line with previous studies~\cite{gauvin2020gender}.

A possibly long-lasting change was in tele-medicine and chatbot adoption, with these services increasing sharply (more than 20-fold in the US shortly after the COVID-19 pandemic was declared~\cite{cantor2021and}). Online and phone- or video-call consultations aimed both at limiting exposure and minimising visits. Mobile applications were developed for self-triage, self-scheduling, and as information delivery tools~\cite{ganjali2022clinical}, including prescriptions and test results.


The expectation is that tools such as those described by Shakeri et al. (2021) will not only continue to evolve but will also become integral to everyday healthcare provision~\cite{shakeri2021digital}. However, the sustainable maintenance of these tools and other transformative changes in healthcare need significant restructuring and investment.

\section*{Strengths and Challenges in Digital Epidemiology}

There is a persistent idea that online data and digital data sources are irremediably flawed, when compared with CE studies (bias, privacy concerns, ethical issues, etc.~\cite{salerno2023current, zhao2022biases}).

However, it is increasingly clear that both systems have strengths and weaknesses and that DE can unveil fundamental health-related aspects that traditional methodologies do not provide. It is important to accept that methods to minimize bias in traditional studies are common but all data is imperfect and incomplete~\cite{williams2022data}, and we argue that the main difference is found in the de-biasing process: whereas CE studies consider bias, ethical issues, etc. in study design and data collection, several DE datasets have a broader reach but can only be corrected after collection. 

We summarize some of the strengths and challenges of DE datasets, offer recommendations for the minimization of the latter, and propose future avenues for monitoring and research.

\begin{itemize}
 \vspace{12pt}
	\item \textbf{Complementarity and validation} Different data sources can be useful for syndromic surveillance and they should be used to complement epidemiological datasets by expanding demographics and helping correct bias. Medical tests, epidemiological surveys, etc., should be designed to validate DE datasets, particularly of people with poor internet access or digital literacy, as in~\cite{mcgough2017forecasting}.
    Conversely, people from specific ethnicities have long been subject to discrimination from traditional medical services, and their symptoms might be more fairly evaluated over on-phone appointments; socially anxious individuals might not be visible to sentinel doctors, especially when the symptoms are mild, but appear in online datasets. Moreover, easy and instantaneous access to visual information, such as the one provided by combining multiple digital data streams, allows temporal and spatial analyses, in close to real-time. This facilitates the identification of data gaps, geographical bias, etc., and can play a fundamental role in helping design further epidemiological studies.
    \vspace{12pt}
    \item \textbf{Relation with private sector} Many of the online data collected are from private companies (social media, testing laboratories) and stakeholders such as private hospitals, technology companies, and online platforms will play increasingly key roles in preparedness. Data-sharing practices can be made easier in DE approaches and health systems should increasingly consider the private sector as long-term partners in preparedness rather than only when emergencies are ongoing and privacy risks are higher (see Table~\ref{tab:Table 1}).
    \vspace{12pt}
    \item \textbf{Data availability and structure} The amounts of available data are fast increasing and these offer an immense potential for health-related analysis, in close to real-time. As much as possible, international standards or guidelines should be agreed upon for collecting, analysing, and sharing relevant medical and non-medical datasets, including requiring enough metadata on criteria for inclusion, demographics, limitations, etc., to allow for comparison and possible extrapolation. Governments should provide greater transparency and research access to their datasets, including epidemiological data and risk factors for acquisition, following a FAIR (Findability, Accessibility, Interoperability, and Reuse) framework~\cite{wilkinson2016fair}.
    \vspace{12pt}
    \item \textbf{Disease identification} Even when data is good, classifying a disease remains challenging. For example, different respiratory viral infections often share symptoms and therapeutics and there is little incentive to laboratory-test them. Therefore, novel tools, sensitive to subtle syndromic differences, can aid in disease identification and monitoring, especially as these diseases show disrupted seasonality and gain relevance~\cite{den2022decline, varela2023effects}. Electronic Prescription Records also offer interesting potential to infer incidence and prevalence of diseases, by using drug usage as a proxy for some diagnoses.
    \vspace{12pt}
    \item \textbf{Digital Adaptation} This shift offers great potential in exposure reduction or in reaching neglected populations, but these tools raise important ethical concerns (Table~\ref{tab:Table 1}). Moreover, the groups that could most benefit from free and fast digital tools are also often the ones experiencing higher barriers (during the COVID-19 pandemic underprivileged communities were hit harder), creating an important digital health paradox~\cite{wong2022dawn}. In the case of tele-medicine, investment in computational infrastructure is paramount with good video and sound systems required from both clinical and patient sides. With respect to wearable sensors, these could be combined with other datasets to allow population-based clustering of behavioural risk factors such as physical inactivity or poor diet. As for DCT, the argument that some people using them was better than no-one at all ignores that an app being used by few leads to a false sense of safety, as no alerts are given out. It also ignores that its effect could be so small that it would not compensate the cost of development~\cite{reyna2021virus}, especially if it only benefited already well-off sub-populations. Overall, it is essential to involve representatives of different communities when designing such solutions~\cite{tan2022call} and further research on existing barriers (technological, sociological, operational) and effectiveness should be prioritized.
    \vspace{12pt}
    \item \textbf{Public Participation} Studies involving participatory data collection can not only collect valuable data and improve demographic coverage but also teach data literacy and give the communities control over decisions that affect them~\cite{chan2020putting, de2021relationship}. These can also aim at creating trust relationships, changing power dynamics, fostering debate, and building collaborative knowledge sharing~\cite{williams2022data}. 
    \vspace{12pt}
    \item \textbf{Research} Access to public or private datasets remains challenging and there is poor inter-connection between academia, tech solution providers, and public health implementation. Academics have offered models, machine-learning-based pattern-identification systems and useful datasets~\cite{won2017early, koppeschaar2017influenzanet} but these are rarely used by public institutions and decision-makers~\cite{morgan2022better}. Researchers, companies, and governments should have better incentives to publish the effectiveness of their models and technologies in peer-reviewed journals and through appropriate clinical evaluation, allowing for comparison and identification of best-practices. These should be followed by specific programs allowing for \textit{de facto} implementation of tools deemed more promising.
    \vspace{12pt}
    \item \textbf{Communication and Disinformation} Transparency and data sharing are of utmost importance, but large disinformation campaigns might have profited from the deluge of available data, preliminary results being discussed publicly, and accessible data visualization tools. This spotlighted the importance of balancing fast responses with careful in-depth analysis: public health organizations, platform companies, journalists, and authorities need to collaborate to provide the public with clear messaging~\cite{sachs2022lancet}. Multidisciplinary strategies such as monitoring protocols, through the creation of news stories and rumors databases, will be central to defining adequate and timely communication mechanisms~\cite{briand2021infodemics}.
\end{itemize}
 \vspace{12pt}

\begin{figure}[H]
\centering
\includegraphics[width=\linewidth]{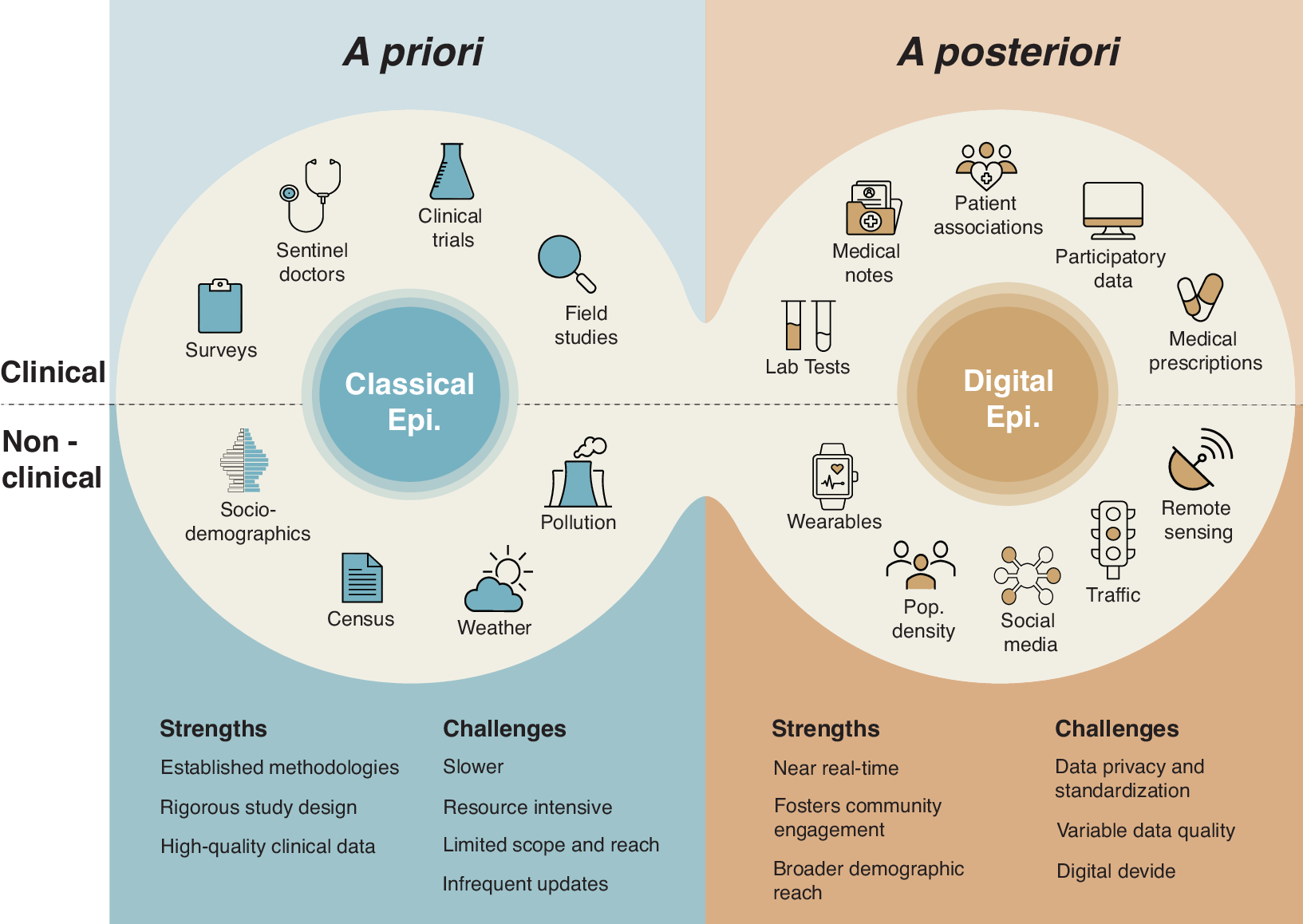}
\vspace{8pt}
\caption{\textbf{Contrasting approaches between Classical and Digital Epidemiology.} This diagram illustrates the distinct methodologies of CE (\textit{A priori}) and DE (\textit{A posteriori}). \textit{A priori} methods, represented on the left, emphasize structured, clinically validated data gathering techniques such as surveys, field studies, and clinical trials, with noted trade-offs in speed and resource allocation. \textit{A posteriori} methods, shown on the right, leverage digital data sources like wearables, social media, and electronic health records, providing benefits in terms of scale and timeliness, yet facing challenges with possible ethical concerns, especially in populations with less digital access (digital divide), or data integrity. Despite their differences, both CE and DE should validate and improve each other, recognizing the strengths and weaknesses of each approach. The ongoing digitization of datasets blurs the boundaries between these categories, highlighting the necessity of a comprehensive and evolving definition for Digital Epidemiology.}
\end{figure}

\begin{table}[p]
\centering
\small
\renewcommand{\arraystretch}{1.3} 
\setstretch{1.0} 
\begin{tabularx}{\textwidth}{@{}p{2.3cm}X X@{}}
\toprule
\textbf{Potential risks} & \textbf{Description} & \textbf{Potential solutions} \\
\midrule
\textbf{Availability bias} & Choosing projects or deciding what questions to answer as a function of popularity or readily available data & Multidisciplinary teams and institutions; include different stakeholders in study design \\
\addlinespace
\textbf{Privacy} & Health-relevant data is considered sensitive by the GDPR, can have different origins, granularity (time, detail, nature), and is often collected without users realizing it & Sound Ethical boards; Technical solutions such as vertical data access systems to guarantee faster access to public and privately-controlled data ~\cite{SPAC} and Federated Learning (FL) to enable models to be trained collaboratively~\cite{rieke2020future,dayan2021federated} without the need for patient data to cross-institutional firewalls. \\
\addlinespace
\textbf{Unforeseen risks} & Biased data collection and predictive and other statistical models, might increase profiling, widen inequalities, and create digital health paradoxes. This is particularly relevant for at-risk populations, such as mental health patients and migrants. & All projects must be weighed in light of the Precautionary Principle. If not possible to determine the risks and harms of collection or analysis, usefulness should be re-evaluated. Sound ethical boards; Transparency~\cite{martuzzi2007precautionary} \\
\addlinespace
\textbf{Self-selection bias} & Individuals select themselves into a group. People who are more comfortable with technology, younger, and from privileged socioeconomic backgrounds will be over-represented in online or participatory studies & Increase digital literacy; understand the bias of the samples or cohorts; use complementary approaches; statistically correct for bias \textit{a posteriori}~\cite{ferretti2022shadow} \\
\addlinespace
\textbf{Repre\-senta\-tiveness}& Specific groups might be over or under-represented in epidemiological datasets and models (e.g. elders might be less present in online or mobile data, but over-sampled in Sentinel Doctor's records; citizens from the Global South might be over-represented in privacy-invading datasets, e.g. individual financial or mobility reports, but under-represented in study design) & Data Auditing and Debiasing, Model Validation, sound Ethical reviews, Transparency, Multi-stakeholder participation; Interaction with communities and decision-makers; "opt-in" strategies to help create public trust and engagement~\cite{mcdonald2016ebola, williams2022data, kostkova2018disease} \\
\addlinespace
\textbf{Data deprivation} & Poor quality or missing data from specific communities, particularly validation data in developing countries, or lack of access to relevant data (e.g. when it belongs to private companies or others unwilling to share) & Improve data-sharing mechanisms (FL, Vertical Systems, see above), International data-sharing templates, policies and regulations; Alternative and independent data collection mechanisms \\
\addlinespace
\textbf{Confirmation bias} & ML and other statistical models often do not require explicit hypothesis testing and offer predictions or results that are statistically significant (due to large datasets) but difficult to validate & External Validation; Complementary methods and datasets; Transparency; Best practices for publishing models and articles \\
\bottomrule
\end{tabularx}
\caption{Biases and Risks}
\label{tab:Table 1}
\end{table}
\FloatBarrier

\section*{Discussion}

Data collection and analysis systems are inherently flawed, but using our definition of DE (utilizing health-relevant data with diverse statistical nature), the main challenge lies in their integration. The desired complementarity will be achieved when datasets designed with \textit{a priori} statistical rigor (e.g., epidemiological surveys, census, weather) are combined with systems designed for different goals (e.g., insurance claims, telephone helplines, social media posts) that require \textit{a posteriori} debiasing. There is no one-size-fits-all method; addressing biases requires recognition and adjustment on a case-by-case basis (see Table~\ref{tab:Table 1}). Multidimensional data visualization is crucial for identifying gaps and patterns, while incorporating behavioral sciences helps interpret seemingly equally relevant data streams and outputs.

Three key steps are necessary: 1) recognizing that CE and DE can validate and improve each other; 2) acknowledging that different biases demand engagement with various communities and stakeholders, beyond statistical and technical solutions; and 3) establishing multidisciplinary approaches and infrastructures. We briefly discuss each below. 

Bias and complementarity are evident in the use of online data and machine-learning models for disease monitoring. An essential concern arises from the amplification of existing inequalities, where underrepresented groups in datasets are also underrepresented in the analysis—applicable to both DE and CE datasets. The term "amplify" carries two meanings: 1) these tools may exacerbate social divides, but 2) they also shed light on these inequalities, making them more visible. While biases in the data mirror real-world disparities, increased visibility might aid in addressing these issues more effectively. Importantly, current debiasing approaches (in both CE and DE, whether they correct datasets or methods) rely on already knowing what the putative biases are: unknown biases are destined to be uncorrected, highlighting the importance of including many demographics and cross-validating tools.
Collaboration plays a fundamental role, requiring long-term projects focused on data quality~\cite{chafetz2022data4covid19} and design testing strategies that prioritize validation rather than just therapeutics.

The challenge of discerning signal from noise~\cite{bento2020evidence} offers another example of the importance of multi-modal approaches: as different motivations drive online searches or social media posts, (made evident in the GFT case), identifying the relevant signal is crucial and we need behavioral or other models, capable of distinguishing between searches driven by actual cases and those influenced by media or fear.

Recognizing the influence of climate change and urbanization on disease control is also vital. 
With 68\% of the world population projected to live in urban areas by 2050, pandemics originating from wildlife-to-human virus jumps are likely to increase~\cite{mora2022over, carlson2022climate}.

Consequently, spatial and seasonal infection standards will vary, and year-to-year epidemic incidence will fluctuate (as observed for flu and RSV)~\cite{varela2023effects, zipfel2021missing}, impacting vulnerable populations. Therefore, it is fundamental that our health and public health systems collaborate with different stakeholders and adapt to a culture of preparedness~\cite{cooper2019creating, meyer2018taiwan}.

Finally, and despite being common-place to mention that epidemiology of infectious diseases is a complex system, strategies are rarely designed to include integration between data, models, ecosystems, and response. Similar to what happens with weather prediction, there was promise of creation of large multidisciplinary institutions (WHO's Hub for Pandemic and Epidemic Intelligence, UK's Center for Pandemic Preparedness, the National Center for Epidemic Forecasting and Outbreak Analytics or the Rockefeller Pandemic Prevention Institute, in the USA) that could also help guide policymakers to implement target measures. However, not only no such institutions are planned for South East Asia, Africa~\cite{WHOafrica} or even Europe~\cite{ECDCstrategy}, it is increasingly clear that these efforts are losing momentum.

Naturally, such approaches require adequate infrastructures, with clear standards, data sharing at local and global scale, and significant funds. But if this is not done now, "when the next major crisis is on our doorstep, we’re not going to be any more prepared to respond to it than we were with this last one"~\cite{RickBright}. 

\section*{Supporting information}

\paragraph*{S1 Appendix.}
\label{S1_Appendix}
{\bf Definitions.} In the Supplementary Material, a comprehensive list of definitions is provided to clarify and standardize the terminology used throughout this paper. Key public health and epidemiological terms such as 'Public Health,' 'Epidemiology,' 'Digital epidemiology,' 'Epidemic intelligence,' 'Surveillance,' 'Tele-medicine,' 'Fore-casting,' and 'Now-casting' are defined to ensure a common understanding.

\paragraph*{S1 Table.}
\label{S1_Table}
{\bf Digital surveillance systems overview.} Presents a summary of the digital surveillance systems evolution for infectious diseases, highlighting the significant changes from pre-COVID-19 methods to adaptations during the pandemic and future perspectives. It covers areas such as syndromic and lab surveillance, contact tracing, digital medicine, spatial analysis, and communication strategies, emphasizing the technological advancements, challenges, and potential improvements for public health management in a post-pandemic world.

\section*{Acknowledgments}
The authors would like to thank Paulo Almeida, Irma Varela, and other members of the Social Physics and Complexity Lab (SPAC-LIP), for discussions and critical reading of the document.

\section*{Funding}
This paper was partially supported by FCT grant DSAIPA/AI/0087/2018 to JGS and by Ph.D. fellowship 2020.10157.BD to SM.

\section*{Author contributions statement}
SM and JGS made the tables and wrote the paper. DP, LP, and JGS supervised and revised the document. 

\section*{Additional information}
The authors declare that the research was conducted in the absence of any commercial or financial relationships that could be construed as a potential conflict of interest.

\end{document}


\maketitle

\section*{Definitions}

\begin{boxD} 
{\bf Public Health} - science and practice of preventing disease, prolonging life and promoting physical and mental health and well-being \cite{winslow1920untilled}\\
{\bf Epidemiology} - study of how disease is distributed in populations and the factors that influence or determine this distribution \cite{EpiGordis}\\
{\bf Digital epidemiology} - originally defined as use of digital data collected for non-epidemiological purposes in epidemiological studies \cite{salathe2012digital}, we offer the alternative use of data collected without \textit{a priori} concerns of statistical representativeness in epidemiological studies\\
{\bf Epidemic intelligence} - all activities related to early identification of potential health hazards, their verification, assessment and investigation aiming to recommend public health control measures \cite{paquet2006epidemic}\\
{\bf Surveillance} - collection, collation, and analysis of data and the dissemination to those who need to know so that an action can result \cite{thacker1983surveillance}.\\
{\bf Tele-medicine} - technology-mediated connection between providers and patients in different locations, allowing for diagnosis, monitoring, triage, patient follow-ups, etc. \cite{kazley2012telemedicine}.\\
{\bf Fore-casting} - predicting future events based on a foreknowledge acquired and implies planning under conditions of uncertainty \cite{lauer2020infectious}.\\
{\bf Now-casting} - prediction of the present, the very near future and the very recent past \cite{giannone2008nowcasting}.
\end{boxD}
\newpage

\section*{Digital surveillance systems overview}

{\footnotesize
\begin{longtable}[c]{
@{}
>{\raggedright}p{0.5cm} 
>{\raggedright}p{1.7cm} 
>{\raggedright}p{2.6cm} 
>{\raggedright}p{2.6cm} 
>{\raggedright}p{2.6cm} 
p{1.2cm} 
@{}}
\caption{Digital surveillance systems}
\label{monitoring}\\
\toprule  
\multicolumn{1}{c}{}&
\multicolumn{1}{c}{}&
\multicolumn{1}{c}{\textbf{Pre-COVID}}&
\multicolumn{1}{c}{\textbf{Post-COVID-19}}&
\multicolumn{1}{c}{\textbf{Future directions}}&
\multicolumn{1}{c}{\textbf{Some References}}\\ \midrule
\endfirsthead
\caption[]{Digital surveillance systems overview (continued)}\\
 \toprule  
\multicolumn{1}{c}{}&
\multicolumn{1}{c}{}&
\multicolumn{1}{c}{\textbf{Pre-COVID}}&
\multicolumn{1}{c}{\textbf{Post-COVID-19}}&
\multicolumn{1}{c}{\textbf{Future directions}}&
\multicolumn{1}{c}{\textbf{\makecell{Some \\ References}}}\\ \midrule
\endfirsthead

{\small\multirow{10.4}{*}{\makecell {\begin{turn}{90}\bf {Syndromic surveillance}\end{turn}}} & & {\bf Clinical:}
Mostly in-person symptoms collection; Sentinel Doctors; Inability to detect outbreaks in real-time; {\bf Online:} Efforts to analyse social media posts and online searches (e.g. Google Flu Trends); Participatory syndromic surveillance (e.g. InfluenzaNet)  & {\bf Clinical:}
Semi-automatic syndromic-based triage; Tele-medicine; {\bf Online:} Increased participatory syndromic surveillance; Very intense media reporting affecting datasets; Hard to discriminate symptoms of different respiratory viruses;  Tools tested in large scale with promising results; Fragmented approaches & Multi-system approach: self-reporting, social media \& traditional methods; systematic testing of different respiratory viruses for validation; improve controls to/and reduce sampling bias & \cite{smolinski2015flu}, \cite{desjardins2020syndromic}, \cite{maharaj2021effect},  \cite{budd2020digital}\\
\addlinespace
\bottomrule
\addlinespace
{\small \multirow{20.4}{*}{\makecell{{\begin{turn}{90}\bf Lab Surveillance}\end{turn}}}
& {\bf Testing} & Lab testing (mainly PCR) of known pathogens; Testing of symptomatic patients or when therapeutic was in doubt; Sentinel Doctors; Highly trained staff and facilities; Poor integration with electronic medical records; Significant delays between sampling, testing and result sharing 
& First pandemic with population-scale screening available; point-of-care (POC) widespread tests and part of the routine (home, local clinic, workplace, etc.); collaboration across commercial, clinical, government, and research organizations; Increased genomic surveilance; cost \& effectiveness of mass testing still debated 
& Adoption of wearables; Novel low-cost methods (molecular, AI/Machine Learning inference); Less invasive sampling techniques (e.g., saliva, cough, breathing \& voice); reinforce communication on testing techniques limitations; simultaneous and automated reporting of test results to patients and clinical \& public-health systems & \cite{han2022sounds}, \cite{tran2022innovations}, \cite{budd2020digital} \\
\addlinespace
\addlinespace
&
{\bf Wastewater surveillance} 
&  History of tracking infectious diseases (e.g. poliovirus), antibiotic use and drug consumption; not integrated with other data, tools and reporting systems (incidence, hospitalizations, etc.) & Used to predict community surges but integration with reporting systems still limited; variants of concern (VOC) identified weeks prior to clinical samples; lack of standardization; lack of granularity (cannot identify outbreak location when area served is too wide) & Expand to other pathogens; invest in standardization; integrate with environmental and veterinary monitoring system to identify diseases with zoonotic potential & \cite{bivins2020wastewater, fahrenfeld2016emerging, castiglioni2015novel, servetas2022standards, kilaru2021wastewater, gomes2022public} \\
\bottomrule

{\small \multirow{10}{*}{\makecell {\begin{turn}{90}\bf {Contact Tracing}\end{turn}}} &
& {\bf Manual:} Well documented method, mainly manual; Sensitive to: community trust, privacy concerns, possible political interference, availability of trained professionals, incomplete databases; {\bf Digital:} Initial efforts to use digital badges and other technological tools 
& {\bf Manual:} traditional method kept with human resource limitations; {\bf Digital:} Deployment of cell-phone based apps; Google \& Apple collaboration; Effectiveness, ethical considerations, security, technical issues raised 
& Improve implementation (technological, social, ethical); Ensure transparency; Include communities and multiple stakeholders; take human behavior into consideration  & \cite{muller2021contact}, \cite{olu2016contact}, \cite{akinbi2021contact}, \cite{danquah2019use} \\
\addlinespace
\bottomrule
\addlinespace
{\small \multirow{15}{*}{\makecell {\begin{turn}{90}\bf {Digital Medicine}\end{turn}}} & & {\bf Wearables/apps:} medical apps started being prescribed by physicians (e.g. for glucose monitoring); lack of evaluation in Randomized Controlled Trials; low quality of the evidence; {\bf Tele-medicine:} Online and on-phone triage systems; increased number of digital medical records and health-related platforms; {\bf Sensors:} implemented mostly inside hospitals and clinics to measure vital signs & {\bf Wearables/apps:} preliminary results for SARS-CoV-2 infection detection through heart rate variability, oxygen saturation and respiration rate; validation and accuracy to be determined; {\bf Tele-medicine:} digital health solutions became a necessity (tele-medicine, triage for SARS-Cov2 testing or isolation); absence or insuficient regulation; uneven adoption and access: higher-income earners, highly educated, and with chronic conditions being the likeliest adopters & More robust RCTs; invest in coherent and accessible data infrastructures; invest in digital transformation in clinical contexts; support citizen access to internet and mobile interfaces; invest in education and digital literacy; invest in secure and privacy protecting platforms.
& \cite{byambasuren2018prescribable}, \cite{iqbal2022regulatory}, \cite{rockhealth} \\
\addlinespace
\bottomrule
\addlinespace
{\small \multirow{10}{*}{\makecell {\begin{turn}{90}\bf {Spacial Analysis}\end{turn}}} 
& {\bf Mobility} & Mobile phone data used to trace movement across large numbers of individuals; Combined with other information (climatic, vector information, demographics) improved disease spreading predictions (2011 dengue in Pakistan, 2017 chikungunya in Bangladesh) & Facebook's Data-For-Good, Google's COVID-19 Community Mobility Reports; Apple's COVID‑19 mobility datasets; Control measures adapted after inequities were revealed through mobility analysis; phylogeographic component of SARS-CoV-2 genomic reconstruction informed by Google mobility data; limited metadata; no long-term commitment by companies to share data & Evaluate implemented mobility-control policies (timing, effectiveness, and stringency); enrich mobility source data to ensure data accuracy and less dependency on providers; improve data ownership by citizens and consider the need of informed consent prior to sharing (even when aggregated) & \cite{wesolowski2016connecting},\cite{baker2022infectious}, \cite{wesolowski2015impact}, \cite{mahmud2021megacities}, \cite{chang2021mobility}, \cite{zhang2022human} \\
& 
{\bf Geographical Information Systems (GIS)} & Widely used in resource-constrained settings and countries with high burden of infectious diseases; Technological developments and availability of GIS allowed advances 
& Used to manage lockdowns, detect outbreaks, and clusters of infection; main sources of data: contact tracing data, cell-phones (geolocation, payments, points of connection), flights or social media, socio-economic characteristics of the population and its urban structure, polls, participatory GIS and satellite images; web map viewers highly used to inform public and specialists & Exploring the different dimensions and regional patterns of social determinants of health &  \cite{franch2020spatial}\\
\addlinespace
\bottomrule
\addlinespace
\addlinespace

{\small \multirow{1}{*}{\makecell {\begin{turn}{90}\bf {Communication \& Data Visualization}\end{turn}}}
& 
& {\bf Data Visualization:} Easy access to data visualization software tools; rising popularity of data journalism; easy image sharing on social media {\bf Communication:} very rigid and formal communication channels between public health authorities and general public
& {\bf Data Visualization:} Widespread use of dashboards to monitor (disease, resources) and communicate with policy makers, scientists, healthcare providers, general public; data used mainly from traditional \& official sources; can be used to expose inequities and as bait to spread disinformation; {\bf Communication:} poor communication skills from many researchers and public officials
& Further evaluate how data visualizations influence risk perceptions to improve pandemic communication; use it as a tool to improve the population digital literacy; use it as a validation tool within the target communities of the analysis; research impact(s) and possible mitigation roles on disinformation spread & \cite{dong2022johns, boutron2020covid, kahn2022covic, padilla2022impact}\\
\addlinespace
\addlinespace
\addlinespace
\addlinespace
\addlinespace
\addlinespace
\bottomrule

\end{longtable}
}